# Reversibly tuning the insulating and superconducting state in $K_xFe_{2-y}Se_2$ crystals by post-annealing

Fei Han,[1] Bing Shen,[1] Zhen-Yu Wang,[1] and Hai-Hu Wen[1,2,*]

*1. National Laboratory for Superconductivity, Institute of Physics and Beijing National Laboratory for Condensed Matter Physics, Chinese Academy of Sciences, Beijing 100190, China*

*2. National Laboratory of Solid State Microstructures and Department of Physics, Nanjing University, Nanjing 210093, China*

**Since the discovery of superconductivity[1] at 26 K in oxy-pnictide $LaFeAsO_{1-x}F_x$, enormous interests have been stimulated in the field of condensed matter physics and material sciences. Among the many kind of structures in the iron pnictide superconductors[1-8], FeSe with the PbO structure has received special attention since there is not poisonous pnictogen element in chemical composition and its structure is the simplest one. However, the superconducting transition temperature ($T_c$) in iron chalcogenide compounds is not enhanced as high as other iron pnictide superconductors under ambient pressure until the superconductivity at above 30 K in potassium intercalated iron selenide $K_xFe_{2-y}Se_2$[9] was discovered. The insulating and the superconducting state are both observed in $K_xFe_{2-y}Se_2$ with different stoichiometries and some groups have tuned the system from insulating to superconducting state by varying the ratio of starting materials[10, 11]. The recent data from neutron scattering[12, 13] suggest that the superconductivity may be built upon an ordered state of Fe vacancies as well as the antiferromagnetic state with a very strong ordered magnetic moment 3.4 $\mu_B$.**

**Here we show that the superconductivity can actually be tuned on a single sample directly from an insulating state by post-annealing and fast quenching. Upon waiting for some days at room temperatures, the superconductivity will disappear and the resistivity exhibits an insulating behavior again. The spatial distribution of the compositions of the as-grown sample and the post-annealed-quenched one was analyzed by the Energy Dispersive X-ray Spectrum (EDXS) and found to be very close to each other. Therefore it is tempting to conclude that the superconductivity is achieved when the Fe-vacancies are in a random (disordered) state. Once they arrange in an ordered state by relaxation or slow cooling, the system turns out to be an insulator.**

By using the self-flux method, we successfully grown high-quality single crystals of potassium intercalated iron selenide $K_xFe_{2-y}Se_2$. The detailed growing process for the samples is given in Methods. By electrical resistivity measurements we find the resistivity increases with the temperature decreases (as shown in Figure 1a) and the temperature dependence of resistivity $\rho(T)$ has a thermally activated behavior in low temperature region: $\rho=\rho_0\exp(E_a/k_BT)$, where $\rho_0$ and $k_B$ are a prefactor and the Boltzmann constant, respectively. The activation energy $E_a$ was estimated to be about 1.8 meV. Both the two behaviors have been observed in the $K_xFe_{2-y}Se_2$ samples with high iron vacancies by other groups[10, 11]. And Chen's group have tuned the system from insulating to superconducting state by varying the iron content. Here we report a new tuning method: post-annealing. We sealed the crystals under high vacuum in quartz tubes, firstly. Then we annealed them at 200°C, 300°C, and 400°C for 1 h and fast quenching to room temperature, respectively. By electrical resistivity measurements we find the annealed crystals' transporting characters get changed. The sample annealed at 200°C (Figure 1b) has an insulating behavior from 30 to 300 K and the resistivity drops when the the temperature decreases below 30 K. The droping may be caused by the weak superconductivity. The sample annealed at 300°C (Figure 1c) has been tuned to superconducting state and there is a hump-like anomaly at 155 K in the curve of $\rho(T)$. The sample annealed at 400°C (Figure 1d) is also superconducting and the hump-like anomaly shifts to above 250 K. The hump-like anomaly may be a metal-insulator transition since $\rho(T)$ exhibits a metallic behavior below the temperature of "hump". We find that the absolute value of resistivity decreases with the annealing temperature increases from 200 to 400 °C, which means metallic behavior and insulating behavior compete with each other. With the annealing temperature increases the metallic behavior takes the lead.

In Figure 2 we show the temperate dependence of dc magnetization for the samples not annealed and annealed at 200°C, 300°C, and 400°C for 1 h, respectively. The measurement was carried out under a magnetic field of 50 Oe in zero-fieldcooled (ZFC) and field-cooled (FC) processes. Paramagnetic signal is observed in the sample not annealed, and there is not diamagnetization even in the low temperature regime. Very weak diamagnetic signal, which is corresponding to the droping in the resistivity curve, appears below about 26 K in the sample annealed at 200°C. When the annealing temperature increases to above 300 °C, strong diamagnetic signals appear in the samples. We find that the annealing temperature has an important influence on the diamagnetization signal and we get a function of the diamagnetization signal versus the annealing temperature (as shown in the inset of Figure 2). In sharp contrast to it, we find the transition temperature does no depend on the annealing temperature since the diamagnetization signals all appear below 26-27 K in the samples annealed at 200°C, 300°C, and 400°C.

To investigate what effect the annealing process have on the $K_xFe_{2-y}Se_2$ crystals and why the superconductivity appears, we carried out X-ray diffraction on the several crystals. As shown in Figure 3, the peaks from the 00*l* reflections are very sharp, indicating excellent crystalline quality. However, after calculating the lattice parameters of *c* axis, we hardly get any annealing-temperature-dependent variation among the several crystals. The same 00*l* reflections and the close lattice parameter of *c* axis indicate that there is not a structure-transition between the crystals not annealed and annealed. To further to solve the problem why the superconductivity appears, we use the scanning electron microscope and the Energy Dispersive X-ray Spectrum (EDXS to analyse the several crystals. In the the scanning electron microscopic pictures, we observe lots of granules on the surface of the crystal annealed at 400°C (as shown in Figure 4b) while hardly any granules on the crystal not annealed (as shown in Figure 4a). We chose two typical granules to do further research. With the Energy Dispersive X-ray Spectrum (EDXS) to analyse the two granules, we get the chemical composition of the two granules (crossed by the straight line as shown in Figure 4b) as $K_{2.10}Fe_{0.93}Se_2$ and $K_{2.68}Fe_{1.01}Se_2$, which are not able to be described in the formula $K_xFe_{2-y}Se_2$ ($x \leq 1$). We also collected the relative contents of K, Fe and Se elements along the straight line crossed the two granules with the Energy Dispersive X-ray Spectrum (EDXS). The K content increases to maximum at the positions of the two granules while it almost remains as a constant (as shown in Figure 4c) at other positions. On the contrary, The Fe and Se contents decrease to minimum at the positions of the two granules while they also remains as a constant (as shown in Figure 4c) at other positions. Taking the average chemical composition is $K_{0.68}Fe_{1.46}Se_2$ of the crystal not annealed as reference, we find the chemical composition of the main surface apart from the granules in the crystal annealed at 400°C is very close to the crystal not annealed (as shown in Figure 4d). It seems that annealing makes potassium separates out from the crystals. However, the separating out does not change the chemical composition of crystals too much. In addition, it is surprising the crystals annealed lost their superconducting character 20 days later, in which period the crystals were always kept in the argon atmosphere. As shown in Figure 5, after 20 days, strong diamagnetization signal has disappeared in the crystal annealed ay 400 °C, and the insulating state comes out again. Combine the appearing and disappearing process of superconductivity, we attempt to conclude that the superconductivity is achieved when the Fe-vacancies are in a random (disordered) state. Once they arrange in an ordered state by relaxation, the system turns out to be an insulator.

In summary, by post-annealing we tuned the $K_xFe_{2-y}Se_2$ crystals from insulating to superconducting state. And we find that the annealing temperature has an important influence on the diamagnetization signal of the crystals while the $T_c$ hardly depend on the annealing temperature. We also find the tuning is reversible, the superconducting state disappears after about 20 days and the insulating state comes out again. So we get a conclusion that the superconductivity in the $K_xFe_{2-y}Se_2$ crystals is achieved when the Fe-vacancies are in a random (disordered) state.

**Methods Summary**

**Sample preparation.** The single crystals of $K_xFe_{2-y}Se_2$ were grown by using the Bridegman method. First FeSe powders were obtained by the chemical reaction method with Fe powders (purity 99.99%), and Se powders(purity 99.95%). Then the starting materials in the ratio of K: FeSe = 0.8: 2 were placed in an alumina crucible and sealed under vacuum in a quartz tube. The contents were then heated to 1030 °C for 3 hours. Subsequently the furnace was cooled down to 730 °C at a rate of 5 °C/h. Then it was cooled down slowly to room temperature. All the weighing, mixing, grounding and pressing procedures were finished in a glove box under argon atmosphere with the moisture and oxygen below 0.1 PPM.

**Measurements.** The DC magnetic measurements were done with a superconducting quantum interference device (Quantum Design, SQUID, MPMS7T). The zero-field-cooled magnetization was measured by cooling the sample at zero field to 2 K, and the data was collected during the warming up process. The resistivity measurements were done with a physical property measurement system (Quantum Design, PPMS9T) with a four-probe technique. The current direction was changed for measuring each point in order to remove the contacting thermal power.

## Acknowledgements

This work was supported by the Natural Science Foundation of China, the Ministry of Science and Technology of China (973 Projects No. 2010CBA00100), and Chinese Academy of Sciences (Project ITSNEM).

## Competing financial interests

The authors declare that they have no competing financial interests.

**Author Contributions** HHW designed and coordinated the major part of the experiment, including the details about the synthesizing and the annealing, analyzed the data and wrote the paper. FH grown the single crystals of $K_xFe_{2-y}Se_2$ and did the experiment on them, BS and ZYW did part of measurements.

**Author information** Correspondence and requests for materials should be addressed to HHW (hhwen@nju.edu.cn)

## Figure Legends

**Figure 1** Temperature dependence of resistivity for the $K_xFe_{2-y}Se_2$ crystals (a) not annealed, annealed at (b) 200°C, (c) 300°C, and (d) 400°C for 1 h, respectively. The inset of (a) indicates $\rho(T)$ has a thermally activated behavior in low temperature region: $\rho=\rho_0\exp(E_a/k_BT)$ for the crystal not annealed.

**Figure 2** Temperate dependence of dc magnetization for the crystals not annealed and annealed at 200°C, 300°C, and 400°C for 1 h, respectively. The measurement was carried out under a magnetic field of 50 Oe in zero-fieldcooled (ZFC) and field-cooled (FC) processes.

**Figure 3** X-ray diffraction scan showing the 00*l* reflections from the basal plane of the $K_xFe_{2-y}Se_2$ crystals not annealed, annealed at 200°C, 300°C, and 400°C for 1 h, respectively.

**Figure 4** (a) The scanning electron microscopic pictures of the crystal not annealed and 4 plots where chemical composition is collected. (b) The scanning electron microscopic pictures of the crystal annealed. Some granules located on the surface of the crystal. The relative contents of K, Fe and Se elements were collected along a straight line crossed the two granules with the Energy Dispersive X-ray Spectrum (EDXS). (c) The intensity of the Energy Dispersive X-ray Spectrum (EDXS) for K, Fe and Se elements along the straight line crossed the two granules. (d) The chemical composition of every plot along the straight line crossed the two granules.

**Figure 5** After 20 days, (a) the temperate dependence of dc magnetization and (c) the resistivity for the crystal which used to be annealed ay 400 °C for 1 h. (b) The temperate dependence of dc magnetization for the crystal before 20 days.

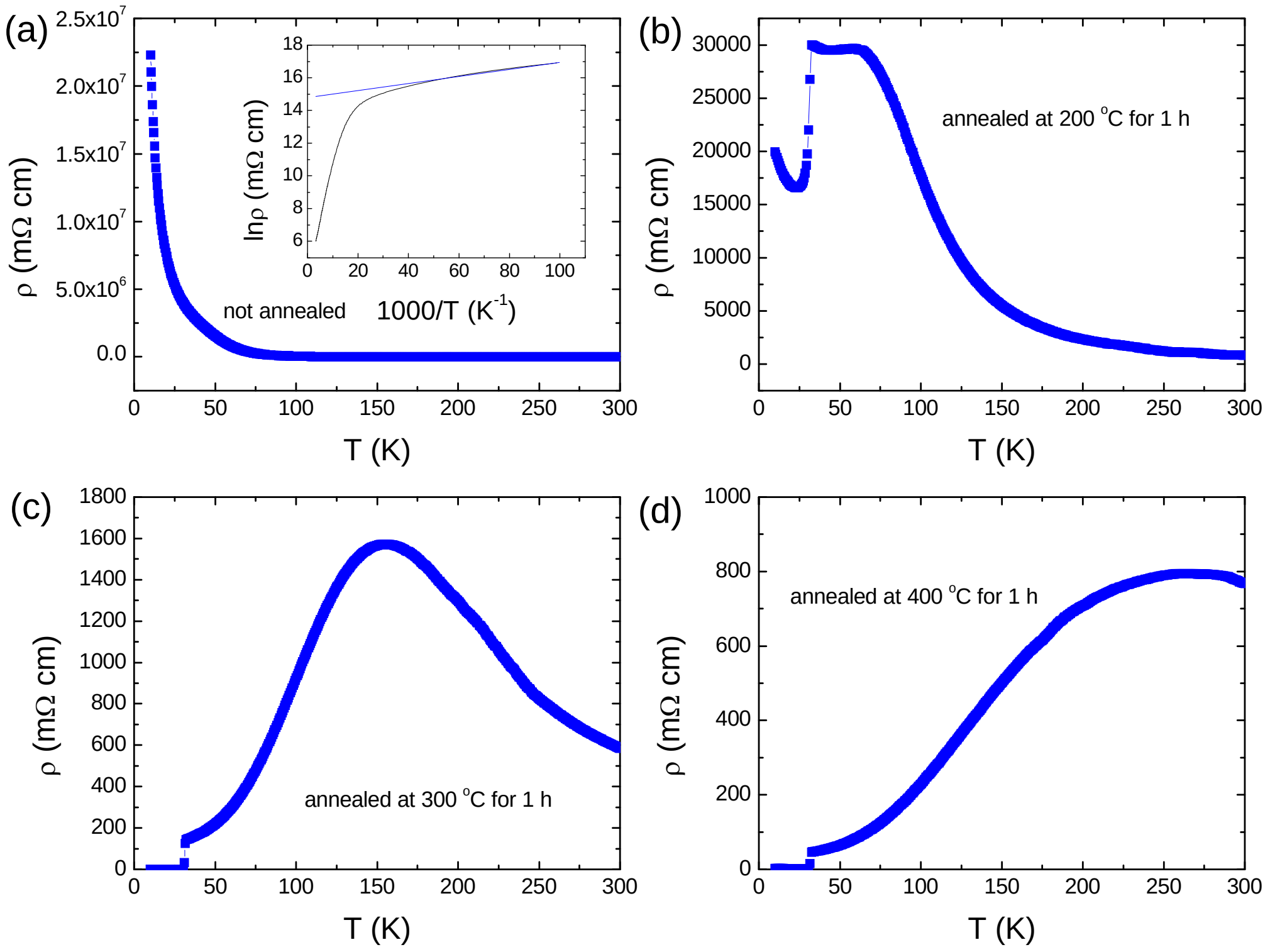

**Figure 1**

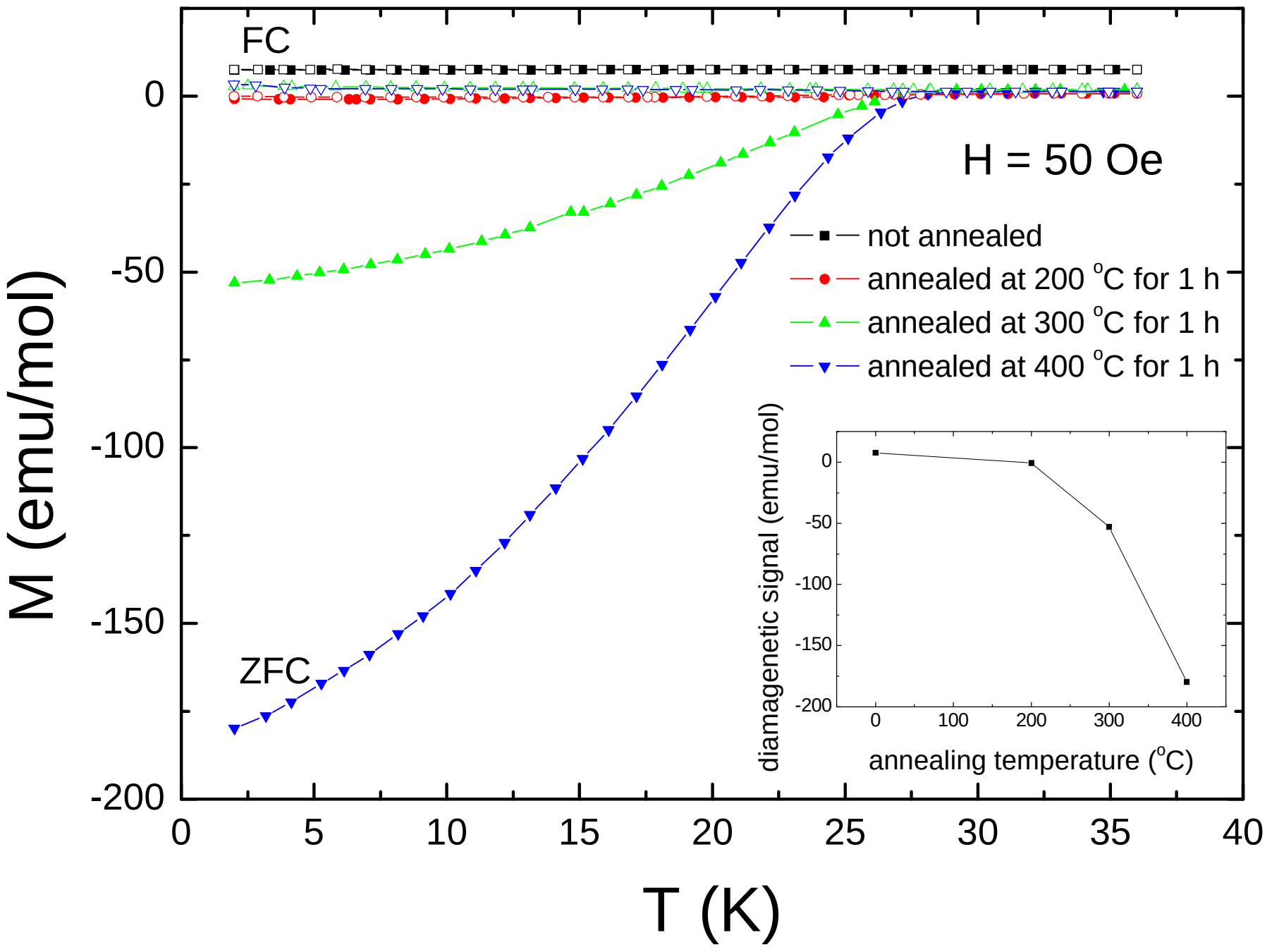

**Figure 2**

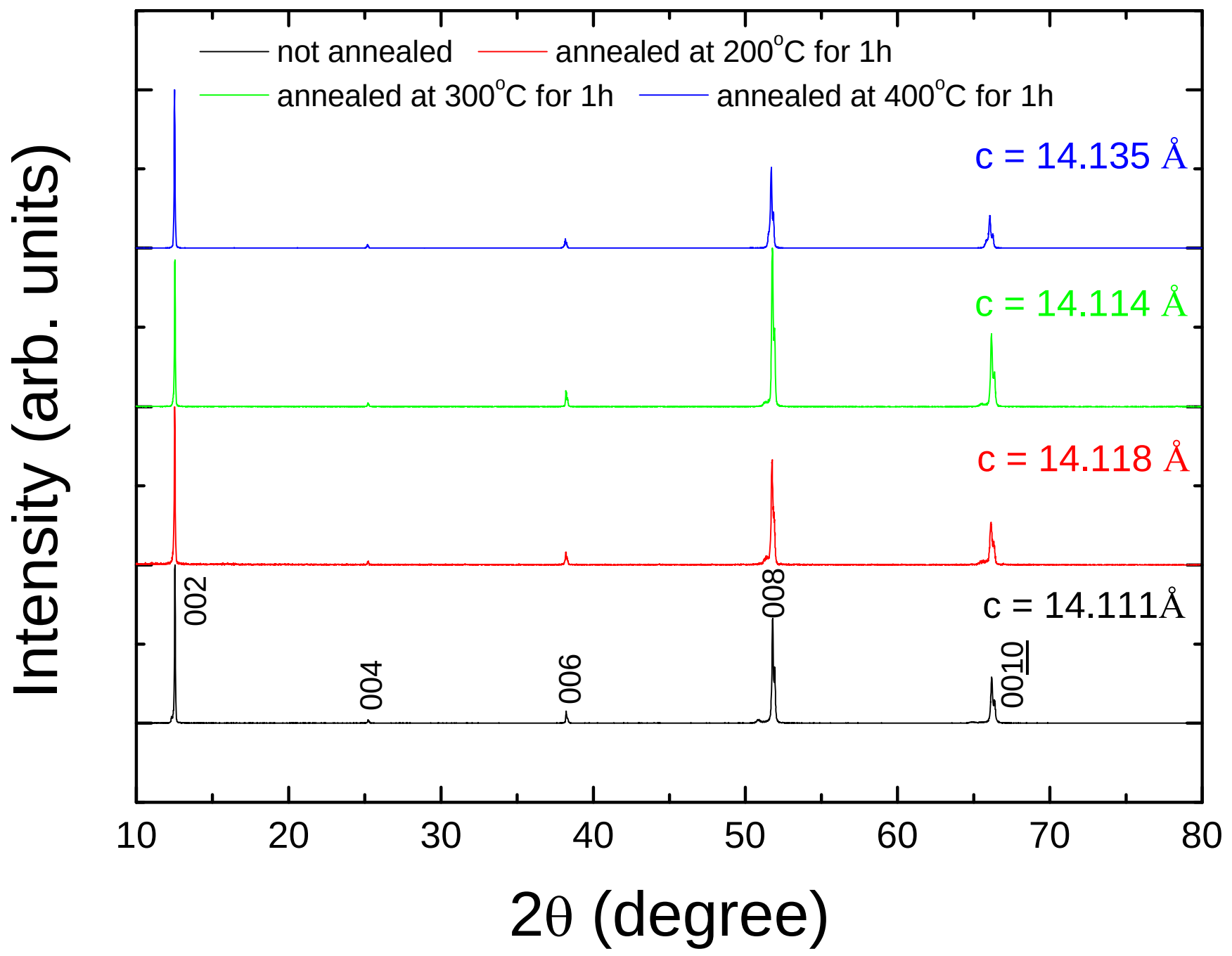

**Figure 3**

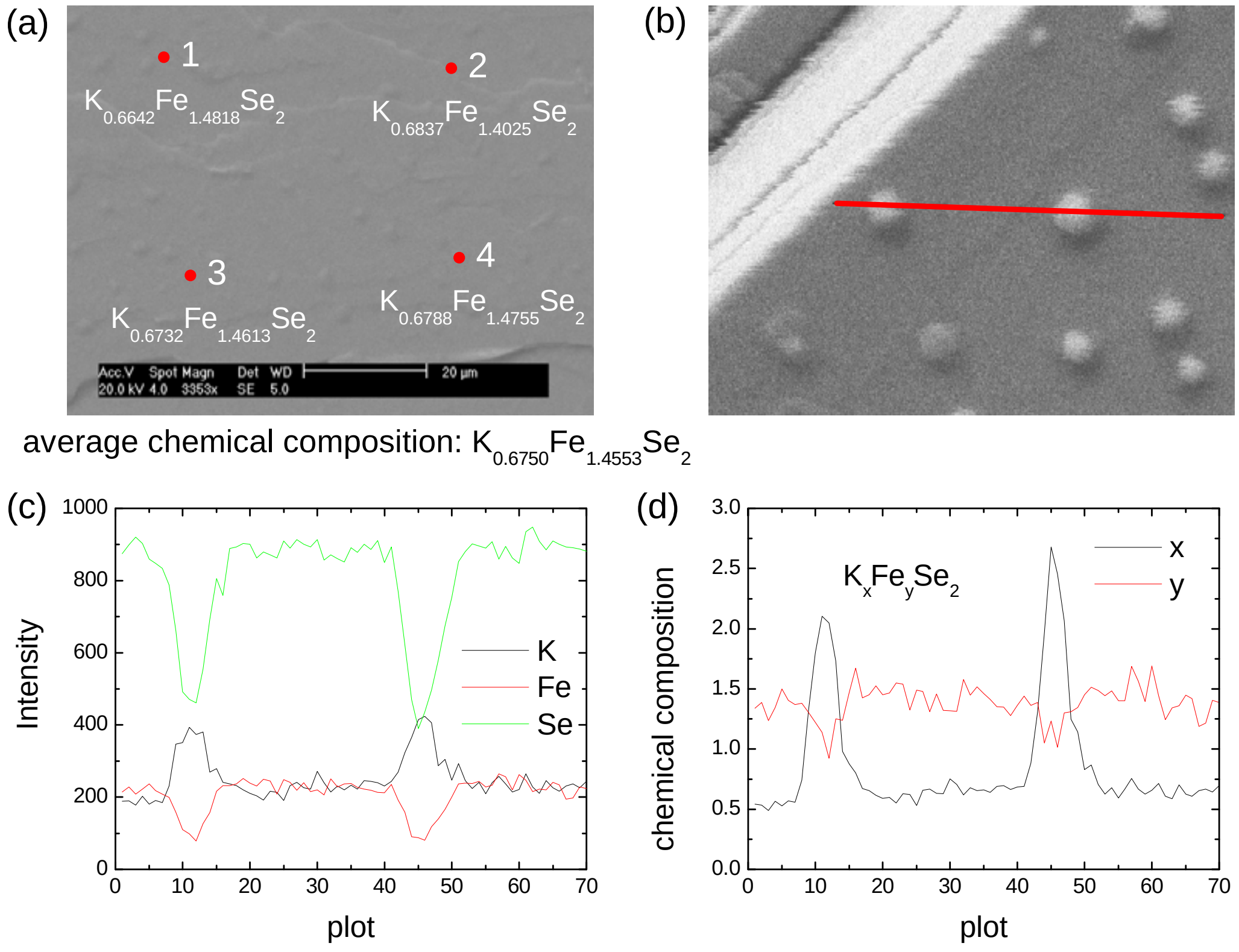

**Figure 4**

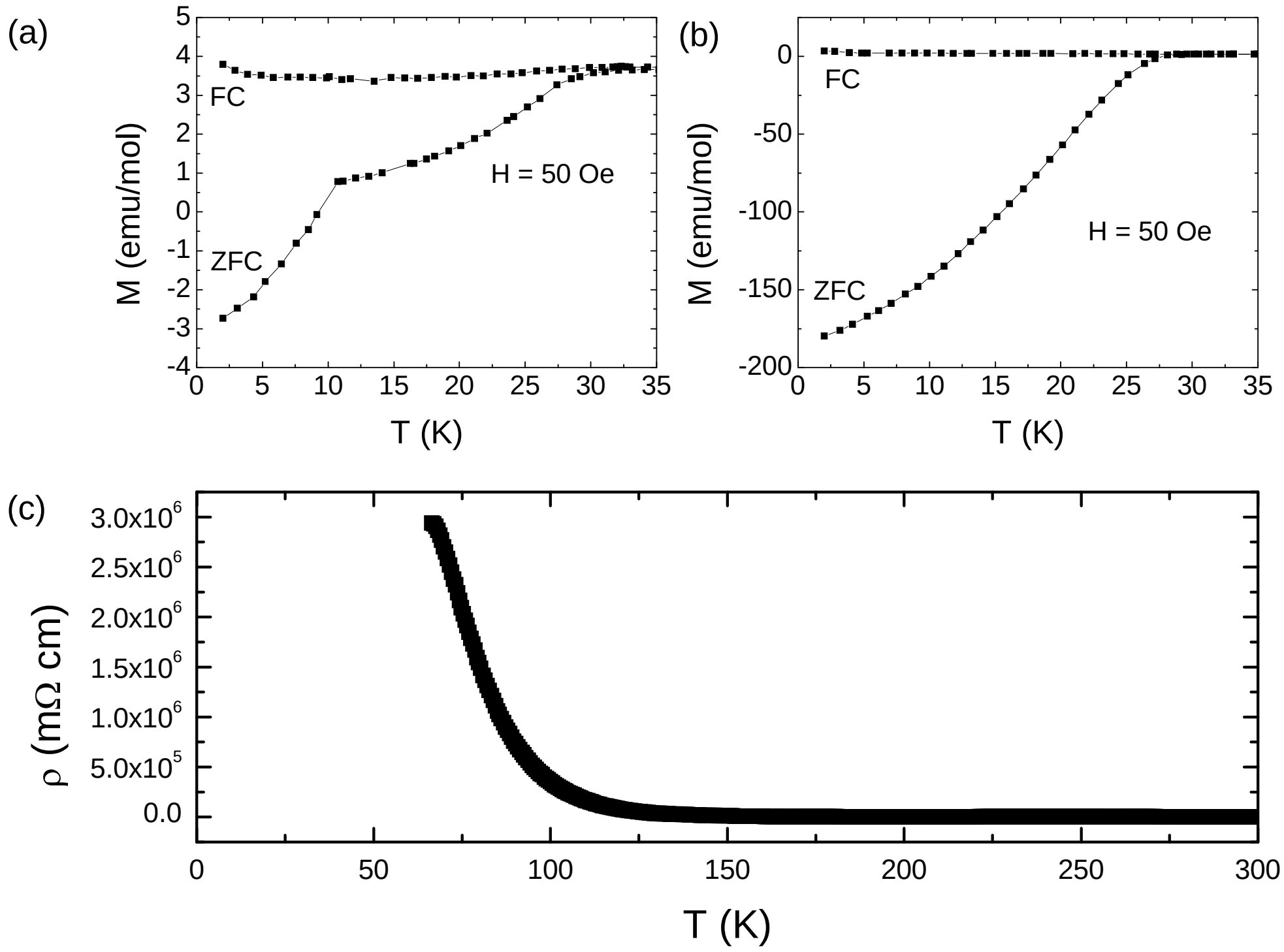

**Figure 5**